\documentclass[conference]{IEEEtran}
\usepackage{cite}
\usepackage{amsmath,amssymb,amsfonts}
\usepackage{algorithmic}
\usepackage{graphicx}
\usepackage{textcomp}
\usepackage{xcolor}
\usepackage{multirow}
\usepackage{booktabs}
\usepackage{bbm}
\usepackage[hyphens]{url}

\def\BibTeX{{\rm B\kern-.05em{\sc i\kern-.025em b}\kern-.08em
    T\kern-.1667em\lower.7ex\hbox{E}\kern-.125emX}}

\title{Towards MoE Deployment: Mitigating Inefficiencies in Mixture-of-Expert (MoE) Inference}
\author{{Haiyang Huang*$^\dagger$\thanks{\noindent\rule{6cm}{0.4pt}}}, 
        {Newsha Ardalani*\thanks{
$^\dagger$ $\ddag$ Work done while interning at Meta}},
        {Anna Sun*},
        {Liu Ke*$\ddag$},
        {Hsien-Hsin S. Lee*},
        {Anjali Sridhar*},
        {Shruti Bhosale*},\\
        {Carole-Jean Wu*},
        {Benjamin Lee*$^\$$}
\\ $^*$Meta AI~~ $^\dagger$Duke University ~~ $^\$$University of Pennsylvania ~~
        $^\ddag$Washington University in St. Louis~~ }
\IEEEoverridecommandlockouts
\begin{document}
\maketitle
\thispagestyle{plain}
\pagestyle{plain}


\begin{abstract}

Mixture-of-Experts (MoE) models have gained popularity in achieving state-of-the-art performance in a wide range of tasks in computer vision and natural language processing. They effectively expand the model capacity while incurring a minimal increase in computation cost during training. However, deploying such models for inference is difficult due to their large size and complex communication pattern. In this work, we provide a characterization of two MoE workloads, namely Language Modeling (LM) and Machine Translation (MT) and identify their sources of inefficiencies at deployment. 

We propose three optimization techniques to mitigate sources of inefficiencies, namely (1) Dynamic gating, (2) Expert Buffering, and (3) Expert load balancing. 
We show that dynamic gating improves maximum throughput by 6.21-11.23$\times$ for LM, 5.75-10.98$\times$ for MT Encoder and 2.58-5.71$\times$ for MT Decoder. 
It also reduces memory usage by up to 1.36$\times$ for LM and up to 1.1$\times$ for MT.
We further propose Expert Buffering, a new caching mechanism that only keeps hot, active experts in GPU memory while buffering the rest in CPU memory. This reduces static memory allocation by up to 1.47$\times$.
We finally propose a load balancing methodology that provides additional scalability to the workload.

\end{abstract}

\section{Introduction}
\label{sec:intro}

The prediction capability of a machine learning model is strongly correlated with the model capacity, (i.e., the number of parameters in the network). In pursuit of accuracy, capacity has grown at an exponential pace of 10 times per year \cite{sevilla2022compute}, accompanied by higher demand for computational resources and extortionate training costs. Sparsely activated neural networks, such as Mixture of Experts (MoE), are attractive model architectures that decouple the requirement for many parameters from the computational costs. In a sparsely activated model, parts of the network are conditionally activated, which reduces training costs. Results from previous works \cite{lepikhin2020gshard, artetxe2021efficient, fedus2021switch, nllb2022, shen2022se, yang2021exploring} show that MoE models reduce training cost yet improve model prediction performance in tasks such as language modeling~\cite{roller2021hash,fedus2021switch,du2021glam,artetxe2021efficient}, machine translation~\cite{nllb2022} and image recognition~\cite{riquelme2021scaling,xue2022go}. While training has been relatively well studied, MoE deployment and inference has received much less attention.

\begin{figure}[tbp]
    \centering
    \includegraphics[width=0.83\linewidth]{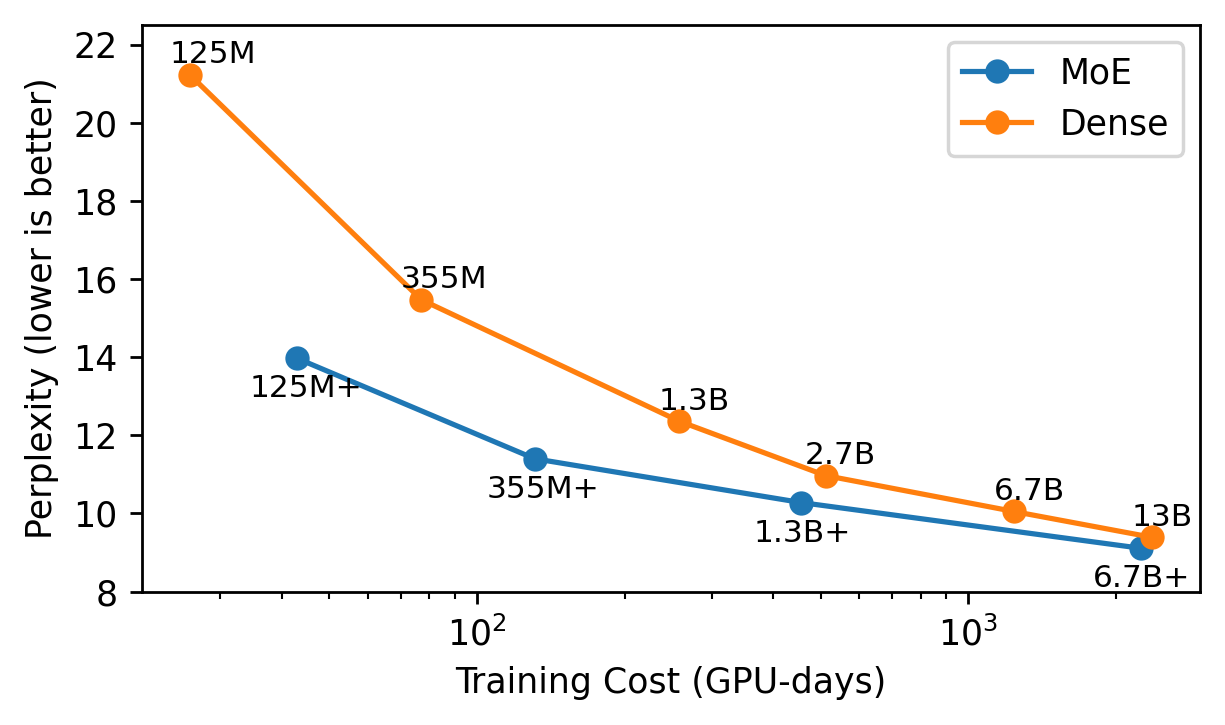}
    \caption{Comparison of MoE and Dense Language Models on training cost and perplexity (the lower perplexity the better in model quality). MoE models can achieve better performance than their dense counterparts at lower training cost
    (Source: Artetxe et. al.~\cite{artetxe2021efficient}).
    }
    \label{fig:moe_dense_train}
\end{figure}

\begin{figure}[tbp]
    \centering
    \includegraphics[width=0.87\linewidth]{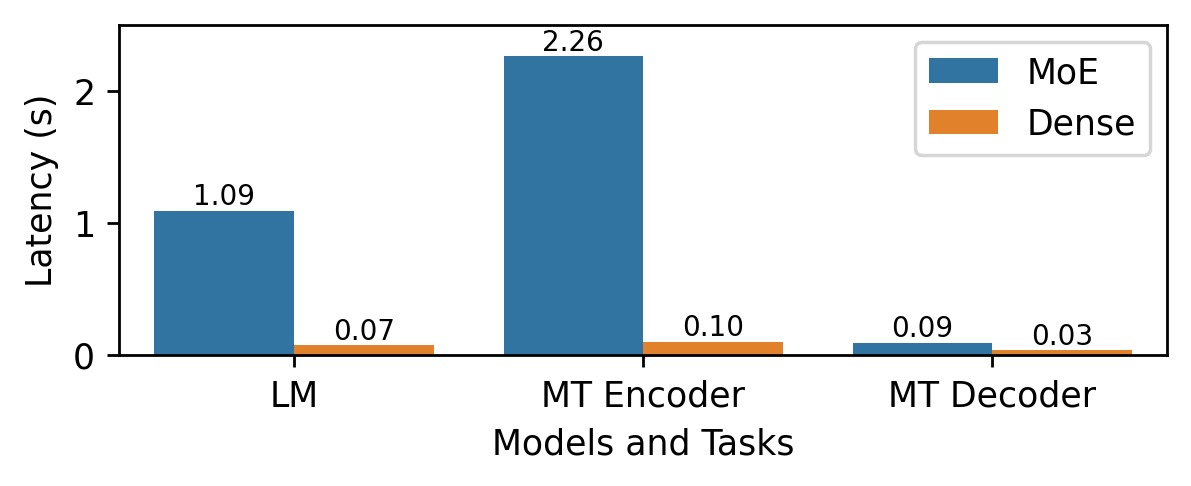}
    \caption{Comparison of MoE and Dense models on single node inference latency. 
    While theoretically MoE models should be able to infer on a similar latency as their flop-equivalent dense counterparts, we find that in practice they are 15$\times$ slower for Language Modeling (LM), 22$\times$ slower for Machine Translation (MT) encoder and 3$\times$ slower for Machine Translation decoder.}
    \label{fig:moe-vs-dense-latency}
\end{figure}

Characterizing and optimizing inference is increasingly important as large language models, like ChatGPT, are deployed for production services.
%
Figures~\ref{fig:moe_dense_train} and~\ref{fig:moe-vs-dense-latency} highlight model prediction capabilities as well as associated training and inference costs between the state-of-the-art MoE and dense model architectures. In Figure~\ref{fig:moe_dense_train}, MoE models achieve the same level of performance and quality (\textit{i.e.}, perplexity) with half of the training cost (GPU-days) compared to their dense counterparts. However, when deployed for inference, MoE models are 15$\times$ slower for language models (LM) and more than 3$\times$ slower for machine translation (MT) compared to their FLOP-equivalent dense counterpart, as shown in Figure~\ref{fig:moe-vs-dense-latency}.

A few strategies have been proposed to reduce MoE inference latency. We might distill MoE models into much smaller dense models with a similar number of FLOPs \cite{fedus2021switch, artetxe2021efficient}. Although distillation reduces model size and inference latency, it also reduces model quality. Lepikhin {\em et. al.} show that a 14.7 billion parameter Switch Transformer model retains only 29\% of its perplexity gain on language modeling after distillation \cite{lepikhin2020gshard}. DeepSpeed-MoE and Tutel\cite{rajbhandari2022deepspeed, tutel} focus on increasing parallelism and optimizing pipelines to increase hardware utilization when deploying MoE models on hundreds of GPUs. These optimizations are scoped narrowly and mitigate inefficiencies in specific kernels for communication collectives and GPU computation. However, these studies lack a comprehensive analysis of inference latency and neglect inefficiencies in the MoE algorithms themselves. 




\textit{In this paper, we provide optimization strategies for efficient MoE deployment, reducing inference costs with minimal impact on model quality.} First, we characterize MoE Transformer deployment on three important axes: inference latency, memory usage, and expert activation. 
Our detailed characterization establishes significant correlations between expert activation patterns and deployment efficiency. Latency and memory usage is high because expert activations are highly sparse and query load is highly imbalanced across experts, 

Second, we analyze unique expert activation patterns to propose a new, optimized gating policy---called Dynamic Gating---and implement it on an open-source, state-of-the-art MoE-based Transformer~\cite{ott2019fairseq}. For Language Modeling (LM) and Machine Translation (MT) across various datasets and subtasks \cite{gao2020pile, nllb2022},
our system prototype for dynamic gating improves inference throughput by 6.21-11.23$\times$ for LM, 5.75-10.98$\times$ for MT Encoder and 2.58-5.71$\times$ for MT Decoder by enabling larger batch sizes and smaller latencies. Our optimization strategies complement previously proposed optimizations on distillation, communication collectives, and GPU kernels. When integrated with other optimizations, our gating policy could achieve even greater benefits. 


Finally, we take a closer look into expert activation patterns, discovering significant imbalance in load distribution across experts but high temporal locality. Based on these two key observations, we propose Expert Buffering, which improves memory efficiency by allocating a fixed, but limited, amount of GPU memory for hot and active experts and relies on CPU memory to buffer all other experts. The less frequently accessed experts are brought into GPU memory as needed, reducing demand for GPU memory significantly. Expert buffering is orthogonal to existing memory management techniques, such as offloading. Our experiments show that expert buffering reduces static memory usage by up to 1.47$\times$ on tasks that demonstrate significant expert sparsity. To balance load, we further propose a priori load balancing based on historical expert activation data, and analyze its benefits for throughput. 


To summarize, our contributions in this paper are as follows:
\begin{itemize}
    \item We provide a thorough characterization of MoE deployment, identifying sources of inefficiencies by breaking down inference latency and memory usage across different components of the model architecture. 
     \item We identify the gating function as a major contributing factor to the high latency and large memory footprint of MoE models. We propose a novel gating policy which significantly reduces latency and memory consumption while also enabling inference with larger batch sizes and a smaller number of GPUs.
    \item We analyze expert activation patterns during inference and discover a significant imbalance in load distribution across experts but high temporal locality.
    \item We propose Expert Buffering, a new caching mechanism that keeps only hot or active experts in GPU memory and buffers the rest in CPU memory. The less frequently accessed experts are brought into GPU memory as needed. This optimization can reduce static memory allocation in GPU by 1.47$\times$.
    \item We propose techniques to balance load across experts to further improve memory usage and system robustness.
    
\end{itemize}

\section{Background}
\label{sec:background}
\begin{figure*}[tbp]
\centering
\includegraphics[width=.95\textwidth]{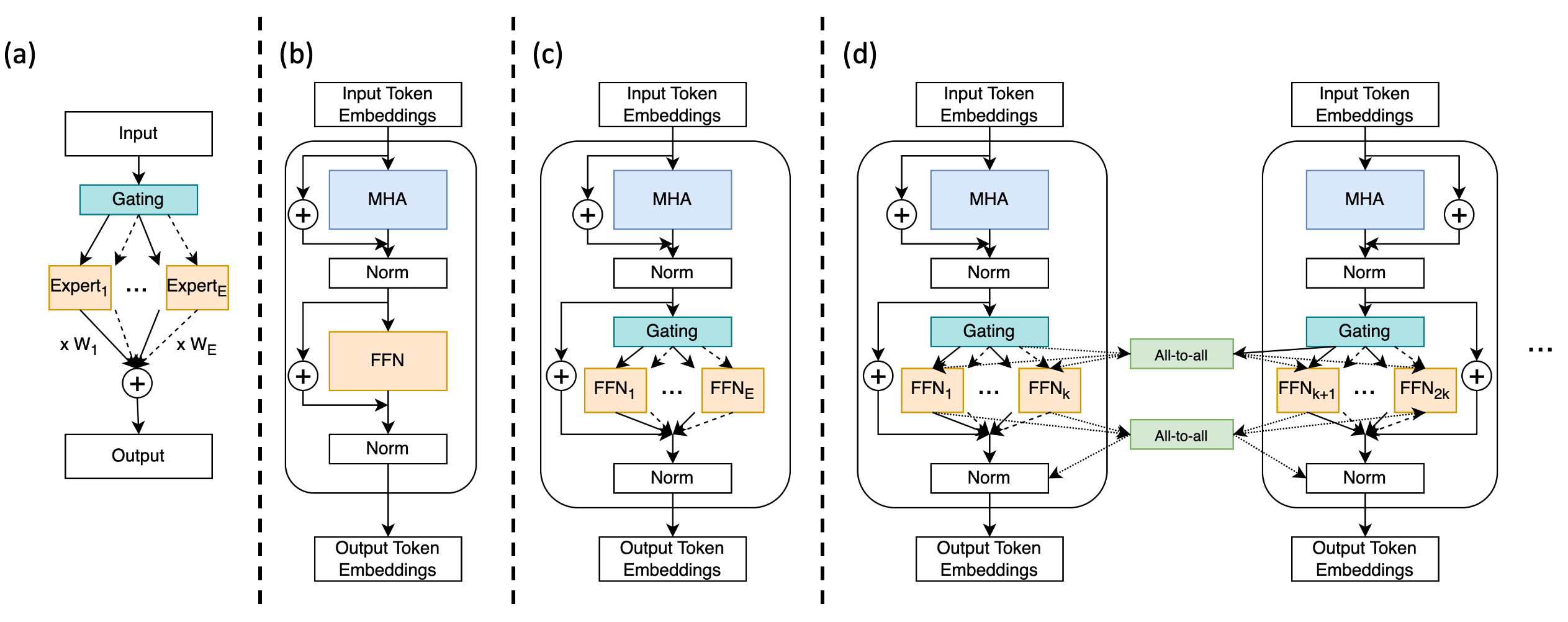}
\caption{Visualization of MoE module, dense transformer encoder layer, MoE transformer encoder layer and MoE transformer encoder layer deployed with expert parallelism. MHA stands for Multi-head attention block, whereas FFN stands for Feed-forward Network block. \textbf{(a)} MoE module introduced in \cite{shazeer2017outrageously} \textbf{(b)} Dense Transformer Encoder Layer. A typical dense transformer layer consists of Multi-head Attention (MHA) followed by an FFN layer. \textbf{(c)} Naive MoE Transformer layer. The single FFN block in dense transformer is replaced by a set of FFNs, called experts, that operate in parallel. Not all tokens are processed by all experts. The gating function decides which experts will receive which tokens.. \textbf{(d)} MoE Transformer with expert parallelism. Each device only holds a subset of all experts. Tokens assigned to non-local expert FFNs are dispatched to their assigned expert via an all-to-all communication collective. 
}.
\label{fig:background}
\end{figure*}

\subsection{Mixture-of-Experts Module}
Using different models for different inputs has long been discussed as a way to improve model versatility and robustness. Mixture-of-Experts (MoE) module \cite{shazeer2017outrageously} is a practical application of this idea for neural networks. An MoE module (Figure~\ref{fig:background}) consists of multiple independent models (called experts), and a gating function that assigns inputs to each of the experts. Each input only activates its assigned expert network, which theoretically allows the model capacity (\textit{i.e.}, the number of parameters in the model) to expand "outrageously" with minimal computation efficiency loss. 

\subsection{Transformer Model Architecture}
The Transformer architecture has gained popularity in computer vision and natural language processing by defining the state-of the-art on multiple tasks in these domains~\cite{vaswani2017attention, dosovitskiy2020image}. From the top down, a Transformer consists of a tokenizer that parses the input into tokens, and an encoder-decoder architecture consisting of dense transformer layers. 
The encoder structure has N dense transformer layers, where N varies from single digits to dozens across different model architectures. Each dense transformer layer is composed of two blocks: a multi-head attention (MHA) block, and a Feed-Forward Network (FFN) block connected by a residual connection, as shown in Figure~\ref{fig:background}. The decoder's structure is very similar to the encoder's, except for an optional MHA layer that attends to encoder output.


\subsection{MoE Transformer Model Architecture}
The MoE Transformer combines the MoE idea with the Transformer architecture. 
In addition to the normal dense Transformer layer, it introduces a new kind of layer with sparse MoEs. Sparse MoE layers replace the FFN block with an MoE block that consists of multiple different expert FFNs. Instead of applying a single FFN to all the input tokens, it first uses a gating function to decide which expert(s) is most suitable for each token, and then routes the tokens to their corresponding expert. Typically, a token is routed to one or two experts in a policy that is referred to as top-1 or top-2 gating. Sparse MoE layers replace the dense transformer layers intermittently in the multi-layer model architecture. 

These modifications grant greater degrees of freedom to the model and effectively expand the model size. Compared to traditional Transformer models, where the FLOP count per batch scales linearly with the number of parameters, MoE networks require much less computation, thus allowing large models to be trained efficiently. 
MoE Transformers have been successful in reducing the training cost of large transformer models\cite{lepikhin2020gshard, fedus2021switch, du2021glam, artetxe2021efficient} and achieving high accuracy in vision, text, speech and multitask learning area\cite{hazimeh2021dselect,roller2021hash, you2021speechmoe, kudugunta2021beyond, gupta2022sparsely}.

\subsection{Expert Parallelism}
Compared to traditional Transformer models of the same model capacity, MoE models offer an interesting trade-off. MoE requires much less computation but much more memory usage. Expert layers deploy many additional FFNs, which increase model size and associated demands for memory capacity near the compute device (\textit{e.g.}, the GPU). To handle this problem, GShard~\cite{lepikhin2020gshard} proposes expert parallelism, which distributes the workload across multiple devices to reduce memory and computation per device. 

With expert parallelism, MoE layers are distributed across multiple devices. Each device holds only a subset of expert FFNs and a copy of all the other parameters. When a token is assigned to experts that reside on other devices, an all-to-all communication collective sends the token to corresponding devices. The tokens are processed by the expert and then sent back by another all-to-all communication. 

At maximum expert parallelism, which allocates one expert per device, memory usage and FLOP count per device are comparable to that from a dense transformer model. Since the gating function is a lightweight linear layer, the overall computational complexity of a batch is about the same as that of a dense transformer with much fewer parameters. Nevertheless, the enormous size, the sparse activation of experts, and the complex communication pattern between devices hosting different experts poses severe challenges during model deployment and inference.



\section{Characterization of the MoE Model}
\label{sec:character}

To characterize the workload of MoE Transformer models, we study two major use cases: Language Modeling and Machine Translation. Language modeling generates the probability an input sequence appears in natural text whereas machine translation maps the input from one language to another. Both tasks are core problems to natural language processing, and are currently major applications of MoE Transformers. 
We choose models in recent publications that achieved state-of-the-art as our testbed. The details of the datasets and models can be found in Table~\ref{tab:expset}.

The MoE model's dense counterparts are selected to be FLOP-equivalent, so they share most of the hyperparameters with the MoE Transformers of interest including hidden dimensions, number of layers and attention heads. The only difference is that the MoE Transformer replaces the FFN layer with an MoE layer every $MF$ layers. Capacity factor $C$, a parameter unique to the MoE Transformer, controls how many tokens can be processed by a single expert. Under the original design, no matter how many tokens are assigned to an expert, the expert will always process a number of tokens equal to $C$ times the sequence length. When too many tokens are assigned to a single expert, excess tokens are dropped and not processed by any expert. When too few tokens are assigned, unused capacity will be filled by zeros. We utilize the capacity factor settings recommended by~\cite{artetxe2021efficient, nllb2022}. Table~\ref{tab:expset} details the experimental setup.

\begin{table}
  \centering
  \begin{tabular}{@{}lcccccc@{}}
    \toprule
    Task & Type & Size & E & MF & CF\\
    \midrule
    \multirow{2}{*}{\textit{LM}} & Dense & 355M & -- & -- & -- \\
     & MoE & 52B & 512 & 2 & 0.05 \\
    \midrule
    \multirow{2}{*}{\textit{MT}} & Dense & 3.3B & -- & -- & --\\
     & MoE & 54.5B & 128 & 4 & 1 \\
    \toprule
    Task & Type & Layers & TD & HD & Vocab\\
    \midrule
    \multirow{2}{*}{\textit{LM}} & Dense & 24 & 1024 & 4096 & 51200 \\
     & MoE & 24 & 1024 & 4096 & 51200 \\
    \midrule
    \multirow{2}{*}{\textit{MT}} & Dense & 48 & 2048 & 8192 & 256206\\
     & MoE & 48 & 2048 & 8192 & 256206 \\

     \toprule
     Platform &\multicolumn{5}{c}{Specification}\\
     \midrule
    \multirow{2}{*}{CPU}
      &\multicolumn{5}{l}{2$\times$Intel Xeon E5-2698 v4 at 2.2GHz }\\
&\multicolumn{5}{l}{with 700GB memory }\\
    \midrule
    \multirow{1}{*}{CPU-GPU}
      &\multicolumn{5}{l}{16GB/s via PCIe 3.0 }\\
    \midrule
     \multirow{3}{*}{GPU}
      &\multicolumn{5}{l}{8$\times$NVIDIA Tesla V100, with 5120 CUDA }\\
       & \multicolumn{5}{l}{cores, 32GB HBM2 memory at 900GB/s }\\
       & \multicolumn{5}{l}{connected by NVLink at 300GB/s}\\

    \bottomrule
  \end{tabular}

  \vspace{0.05in}
  \caption{Experimental Setup. LM: Language Modeling. MT: Machine Translation. E: number of experts. MF: MoE Layer Frequency. CF: Capacity Factor. TD: Token Dimension. HD: Hidden Dimension. Vocab: Vocabulary size.
   E, MF and CF do not apply to dense models.}
  \label{tab:expset}
\end{table}

\subsection{Latency and Memory Consumption}
The most important metrics in machine learning model deployment are execution time (i.e., latency) and memory consumption. A shorter latency ensures a more timely response from the service, whereas lower memory consumption indicates lower resource usage and potential to accommodate larger batch sizes. In this subsection, we compare MoE inference performance along these two axes. The mini batch size is set to 8 for language modeling and 48 for machine translation. We use a dense model of similar FLOPs as the baseline for comparison. 


\textbf{Latency.} Figures~\ref{fig:moe-vs-dense-latency} shows the latency and memory consumption of the MoE Transformers of interest and that of their dense counterparts. Although in theory, MoE Transformers exeucte a similar number of FLOPs compared to the baseline dense models, in practice they are significantly slower. For the Language Model, the dense model requires 74.2ms, whereas the MoE Transformer requires more than 1.09s. For Machine Translation, the dense model executes the encoder and decoder in 101ms and 32ms, respectively, but the MoE Transformer requires 2.26s and 90ms. 

\begin{figure}[tbp]
\centering
\includegraphics[width=\columnwidth]{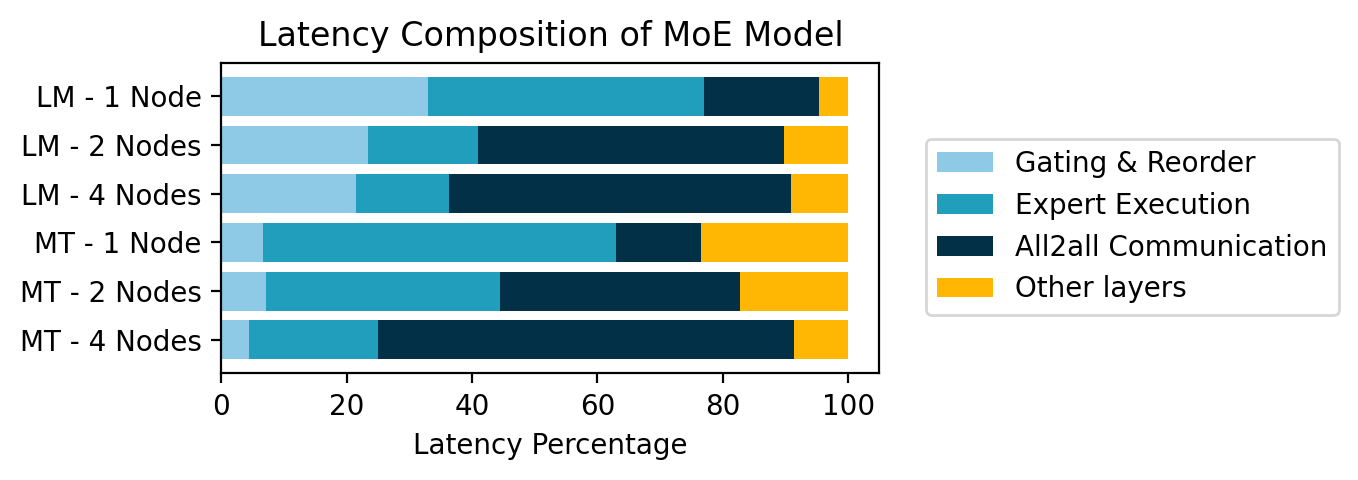}
\caption{MoE Model latency breakdown. Besides all-to-all communication,  other components of the model, such as gating function and expert execution, are also inefficient. Communication overhead increases significantly when more than one node is involved.(Results for batch size=8 for LM and batch size=48 for MT).}
\label{fig:breakdown}
\end{figure}

Figure~\ref{fig:breakdown} breaks down latency under different scenarios.
The latency gap has been previously attributed to the frequent all-to-all communication collective in MoE models.\cite{lepikhin2020gshard} 
While all-to-all collectives does increase latency under multi-node deployment, we note that this is not the only source of latency. 
In Section~\ref{sec:dynamic}, we will discuss these extra sources of latency.
\begin{figure}[tbp]
\centering
\includegraphics[width=.45\textwidth]{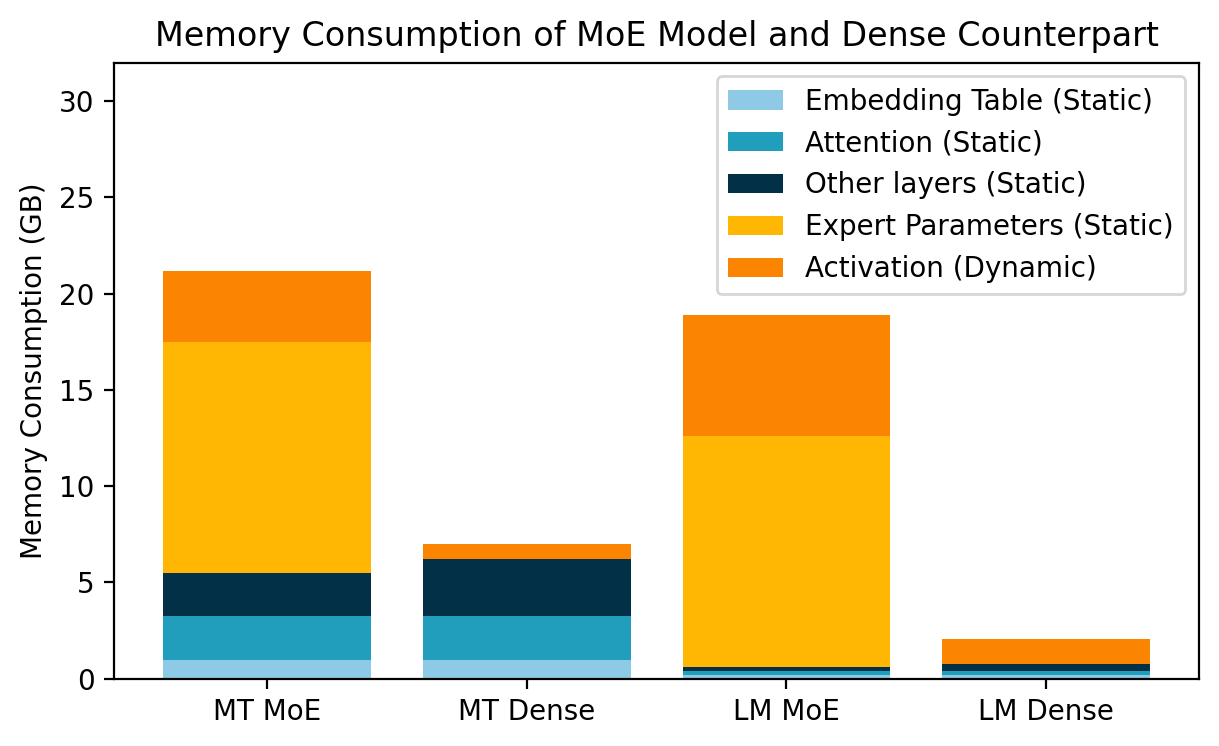}
\caption{MoE vs Dense model memory footprint comparison during inference. The MoE models require significantly more memory usage when deployed on GPUs. Besides the large memory consumption due to the expanded model capacity (introduced by expert parameters), it also requires more memory for activation. (results for batchsize=48 for MT, and batchsize=8 for LM. Note that these are the largest batch sizes that are feasible to run under the baseline implementation.)}
\label{fig:moe-vs-dense-memory}
\end{figure}


\textbf{Memory.} We also observe a large increase in memory consumption for MoE models (see Figure~\ref{fig:moe-vs-dense-memory}). For LM, the dense model only requires 2.2GB on each GPU whereas the MoE model requires 18.88GB at its peak, an increase of 8.58$\times$. For MT, the dense and MoE models use 7.02GB and 21.16GB, respectively, an increase of 3.01$\times$.

We perform a detailed analysis by separating static and dynamic memory usage. Static memory consumption refers to memory allocated to model parameters, whereas dynamic memory consumption refers to memory allocated on demand, usually by network activations. Due to the fact that each GPU accommodates more than one expert during inference, the increase in static memory is expected. However, we observe that the peak dynamic memory consumption also increases significantly in both cases, which is surprising. 

\subsection{Inefficiency of Static Gating} 


What is behind the major overhead in latency and memory consumption? A detailed examination of the latency breakdown sheds light on this matter. While the all-to-all communication collective plays a significant role in multi-node scenarios, other components, such as the gating function and expert execution, also contribute significant inefficiencies. Furthermore, a close analysis of the memory trace indicates that memory allocation occurs during the gating and reordering phases. The source of inefficiency in these components warrants further investigation.

The root cause of performance and resource overheads lies in the static gating policy. Recent implementations~\cite{kim2021scalable,rajbhandari2022deepspeed, roller2021hash, artetxe2021efficient} of MoE Transformer models usually assume the number of tokens assigned to each expert is roughly the same because the loss function during training accounts for load balance. As a result, the token distribution process is simplified to an all-to-all collective that distributes the same number of tokens (see Figure~\ref{fig:dynamic}(a)). 

The Capacity Factor $C$ defines the number of tokens processed by each expert in one batch. If the gating function assigns fewer tokens than an expert's capacity, the rest of the capacity will be filled by placeholders (\textit{i.e.}, zero vectors). If the gating function assigns more tokens than an expert's capacity, excess tokens are dropped by the expert and their information will be retained only by the residual connection. Token drop is undesirable as it harms accuracy. To avoid information loss and accuracy fluctuations, capacity factor $C$ is usually set at high values during inference. While this safeguards accuracy, it increases latency and memory costs.

\textbf{Waste Factors.} For Language Model, where the number of experts $E$ = 512 and the Capacity Factor $C=0.05$, the number of tokens processed by an expert in a sequence $S$ is $ECS$ = 512$\times$0.05$\times S = 25.6S$. The amount of computation the device must actually perform, instead, is only $2S$ since the model implements top-2 gating and each token would be processed by two experts. Therefore, the waste factor is $25.6S / 2S = 12.8$.

For Machine Translation, the analysis is similar. The number of tokens processed by each expert is $ECS = 128 \times 1 \times S= 128S$. However, the amount of computation the device must actually perform is $2S$ as well, which leads to a waste factor of $128S / 2S = 64$. The huge waste factor suggests that typical MoE models perform a large amount of excess computation and communication as well as consume a large amount of extra memory. 

Our question is whether this over-provisioned resource usage is avoidable.
If the workload is well balanced such that token allocations across experts are comparable, resource waste can be reduced by simply scaling down the Capacity Factor. On the other hand, if the expert activation is sparse, scaling down the capacity factor is not an option because doing so increases the chance of dropping tokens and harming model accuracy.

\label{sec:dynamic}
\section{Analysis of Expert Activation Patterns}\label{sec:expass}
\label{subsec:activation}
To understand whether such a huge waste factor is necessary for service stability, we will study expert activation patterns across two applications of MoE models, namely language modeling and machine translation. Moreover, we propose two optimizations (dynamic gating and expert caching) to reduce the waste factor and improve latency and memory consumption.

\subsection{Language Modeling Case}
For Language Modeling, we use the PILE dataset~\cite{gao2020pile} as the input, which is the validation set used in prior work~\cite{artetxe2021efficient}.  We select three domains (Wikipedia, PubMed and Github) from the PILE dataset to study the effect of different input data on the expert activation patterns across time (\textit{i.e.}, consecutive batches).
%

We visualize the results in Figure~\ref{fig:mt_pattern}. Each row represents a batch and each column represents the load of a particular expert. A more intense color indicates the expert receiving a higher portion of all tokens in a batch. As shown in Figure~\ref{fig:mt_pattern}(a), load distribution across experts is highly imbalanced.
There exists multiple hot experts that always get a large share of tokens (multiple lines of intense color), and the other experts consistently receive a small amount of tokens (lines of lighter colors). 

In the most extreme cases, Figure~\ref{fig:mt_sparse} indicates there exist experts that never get any tokens.
Due to the static gating policy, these experts still receive and process empty token placeholders, introducing a huge waste of computational resources.
As shown in Figures~\ref{fig:mt_pattern}(a) and ~\ref{fig:mt_sparse}, the set of hot experts and their hotness level varies across domains even though all domains consistently exhibit a high-degree of sparse expert activation.  

\subsection{Machine Translation Case}
\begin{figure*}[tbp]
\centering
\includegraphics[width=.8\textwidth]{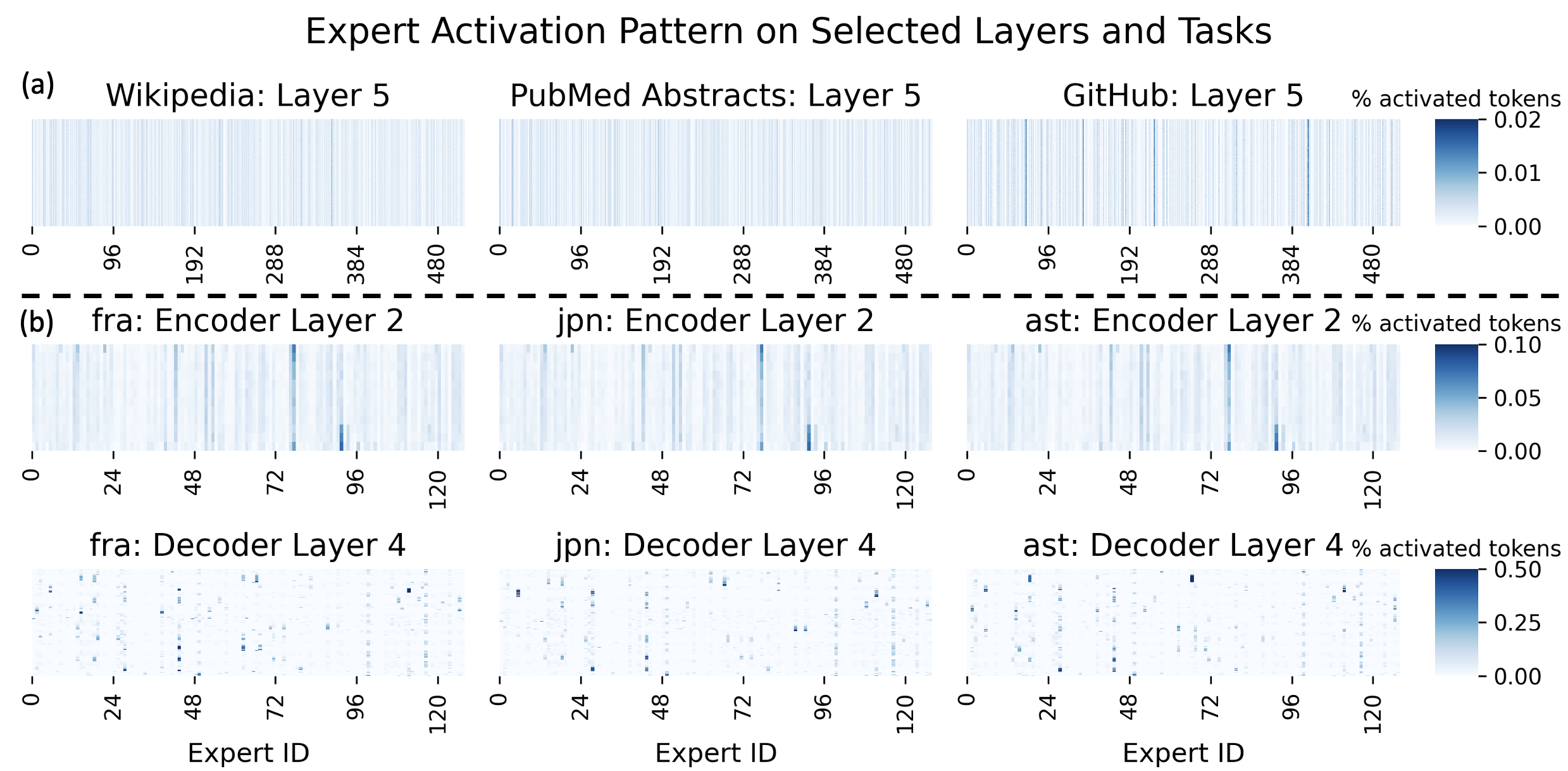}
\caption{Visualization of the expert activation pattern on selected layer of (a) language modeling and (b) machine translation. Activation is normalized. The expert activation pattern exhibits strong imbalance on all the tasks, and the imbalance is consistent. Specifically, on machine translation decoder the sparseness is enormous, and the expert also demonstrates strong temporal correlation.}
\label{fig:mt_pattern}
\end{figure*}

\begin{figure}[tbp]
\centering
\includegraphics[width=.48\textwidth]{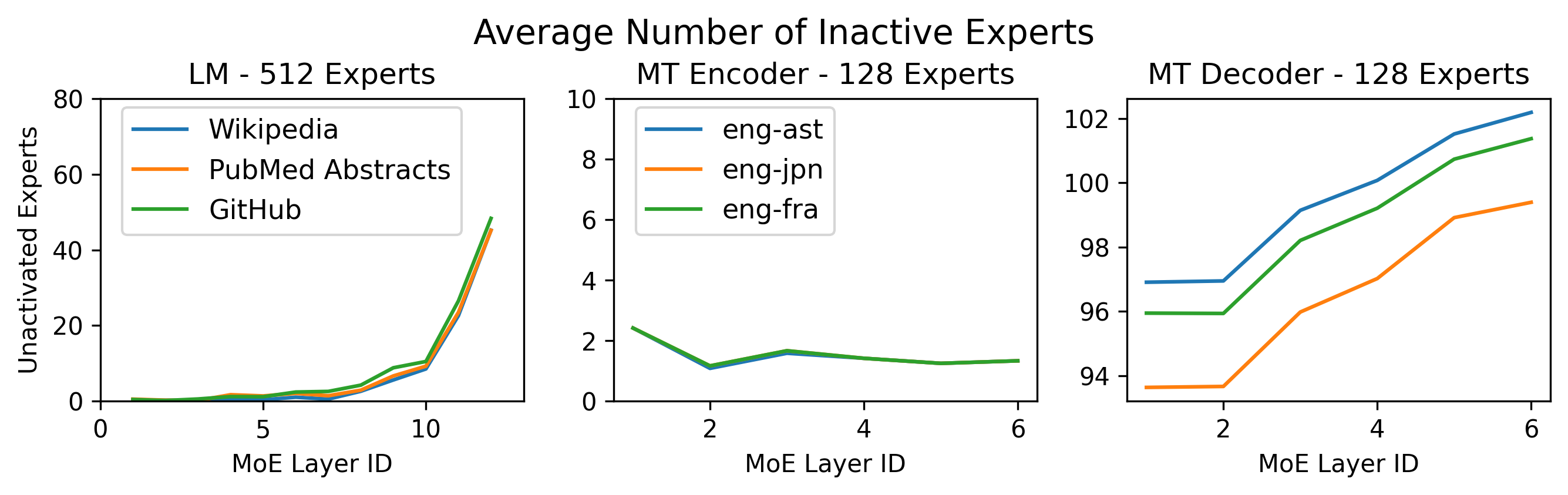}
\caption{Average number of inactive experts on Language Modeling and Machine Translation. Most, if not all experts are activated throughout the LM and MT encoder. However, activation on MT decoder is extremely sparse, even if we utilize a batch size of 96 under dynamic gating policy.}
\label{fig:mt_sparse}
\end{figure}

For Machine Translation (MT), we use the original validation dataset NLLB-200~\cite{nllb2022}. We use English as the source language, and select three different target languages (French, Japanese and Asturian).
Expert activation on MT for randomly selected layers is visualized in Figure~\ref{fig:mt_pattern}(b). 

Machine Translation models also exhibit load imbalance and a small fraction of experts that are more hot than others, and the load imbalance is even more pronounced. Certain experts on both encoder and decoder has received a large share of all tokens that is almost half of the full batch, whereas many experts maintain a low degree of activation.

We further inspect whether expert sparsity exists on the encoder and the decoder of the model. Figure~\ref{fig:mt_sparse} demonstrates the expert sparsity level on the encoder and decoder on all three tasks. We find that the encoder activation is mostly dense,  that most of the experts are activated at all times. The decoder activation is extremely sparse (about 75\%). 



We visualize the selected activation pattern of the encoder and decoder in Figure~\ref{fig:mt_pattern}(b). The activation is normalized within a batch, and the color intensity is a measure of load intensity, representing the percentage of tokens assigned to each expert within a batch.
The detailed activation shows that the expert activation pattern in machine translation is similar across different languages. 
The encoder architecture captures the source language properties which is the same across all three tasks (English). To our surprise, we found that expert activation is more or less similar across different target languages as well as decoder architectures.

A closer inspection on the expert activation on the decoder shows that the expert sparsity has a strong temporal locality. The intense color representing high load of expert usually appears as lines, suggesting that an expert is active across consecutive batches. This implies temporal locality for hot experts. This observation is a key motivation for expert caching discussed in Section~\ref{sec:expcache}.


\section{Dynamic Gating Optimization}
\label{sec:dynamicgate}


\begin{figure}[t]
\centering
\includegraphics[width=.48\textwidth]{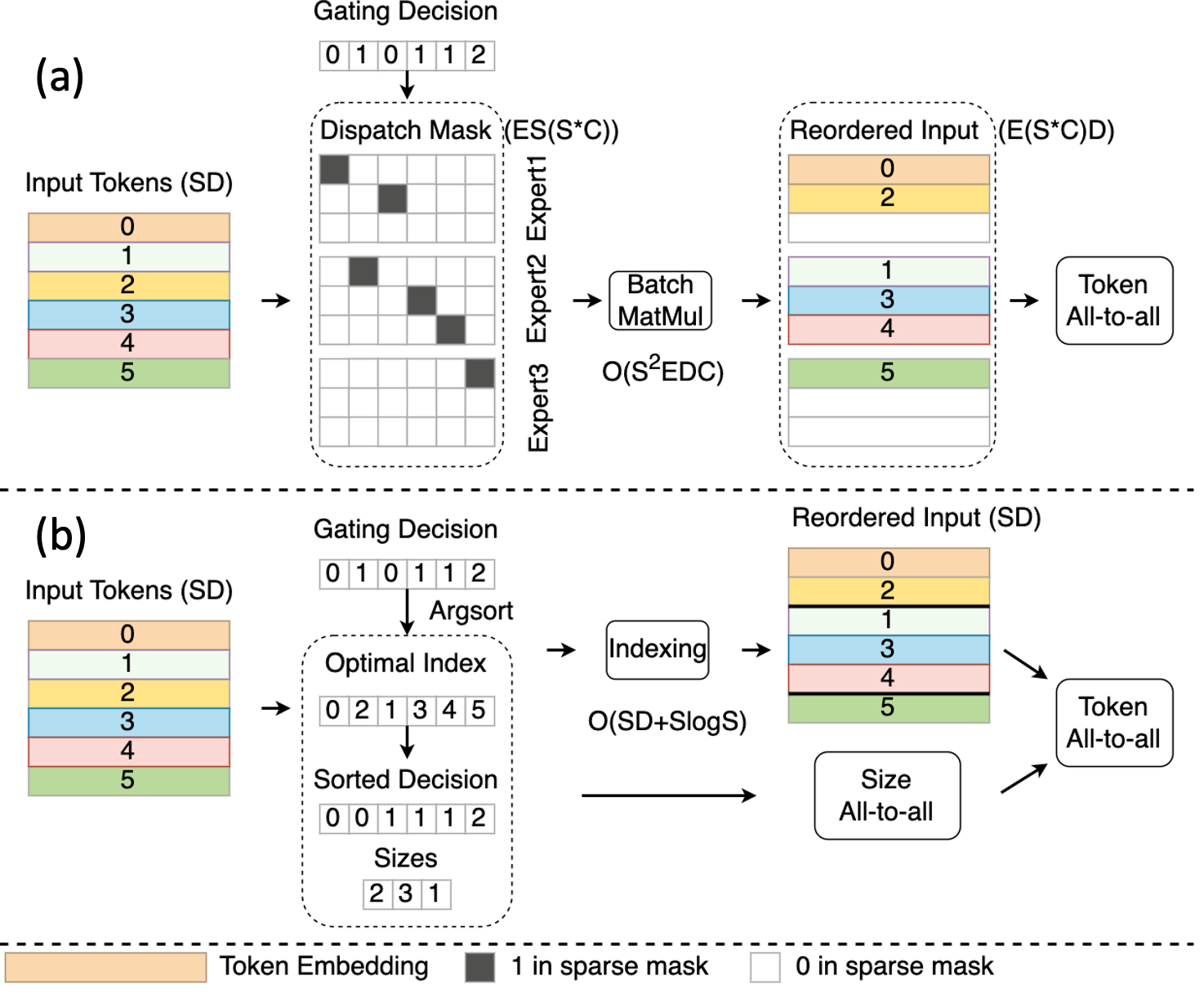}
\caption{Comparison between the static gating in \cite{lepikhin2020gshard,artetxe2021efficient} and our implementation of dynamic gating. For simplicity, we assume E=3, S=6, C=0.5 and top-1 gating in this example. Shapes of tensors are recorded in parentheses. (a) Static Gating. Under a static gating policy, each expert always processes a predefined amount of tokens, which may lead to token overflow or empty tokens. See Section~\ref{sec:dynamic} for details. (b) Dynamic Gating. Under a dynamic gating policy, each expert only processes the tokens that are assigned to it. The token distribution mechanism is simplified with less complexity, and the communication and computation are reduced. }
\label{fig:dynamic}
\end{figure}
The observed activation patterns demonstrate a distinct gap between assumptions in system design and inference performance. Naively increasing expert capacity may still not prevent token overflow for some experts, but will create extra redundancy and waste for other experts. While previous studies also notice the imbalanced activation across experts \cite{tutel, rajbhandari2022deepspeed}, existing solutions retain a static gating policy, which increases $CF$ when severe imbalance appears \cite{tutel}. Our conclusion is that static gating increases resource waste and fixed expert capacity is not the optimal solution for the distribution of tokens to experts. The constraints imposed by static capacity should be removed and the gating function should be dynamic. 

Nevertheless, changing the gating policy to allow dynamic sizes for experts is non-trivial. Major, existing implementations~\cite{tutel, artetxe2021efficient, riquelme2021scaling} do not support dynamic. They rely on static capacity to guarantee that message sizes of all-to-all collectives are the same, which simplifies the communication.

\subsection{Case for Dynamic Gating} 

Figure~\ref{fig:dynamic}(a) visualizes the static gating policy. In this example, we assume a sequence length (total number of tokens in a batch of sentences) $S=6$, number of experts $E=3$, capacity factor $C=0.5$ such that static capacity is $S\times C=3$. We assume top-1 gating such that each token is assigned to only one expert. 

After the gating function generates the gating decision, the static gating policy translates it into $E$ dispatch masks, each of size $(S, (S\times C))$. The dispatch mask is generated as follows. If token $i$ is assigned to expert $e$, then the process will check if the $e$-th mask still has capacity. If so, the $i$-th column of the first empty row will be marked as 1, whereas the other numbers are kept to be 0. This process gives us a sparse dispatch mask that is a tensor of the size $(E,S,(S\times C))$, in which at most $S$ entries are 1's due to potential token dropping. Note that this matrix is highly sparse. The input tokens will be multiplied with the dispatch mask to reorder inputs into $E$ sets of inputs, each with $S\times C$ tokens. Each set of inputs will be sent to its assigned expert's device.


Figure~\ref{fig:dynamic}(b) shows how the gating function must be redesigned when the number of tokens transferred between devices is variable. 
Our implementation simplifies the distribution process. We find an array of indices that can sort the array by performing an argsort. This set of optimal index prepares the order of tokens for dispatch to devices. By counting the number of occurrences of each expert, we know the exact size of the dispatched input to each device. 

Because the sizes of dispatched input are variable \cite{he2021fastmoe}, we adopt a two-step approach. First, we use an all-to-all collective to inform each device the size of the incoming tokens. Second, we use another all-to-all collective to perform the real token transfer. The size all-to-all collective is launched as soon as sizes are known, maximizing overlap with other kernels. Meanwhile, inputs are reordered based on the optimal index for each token and the reordered input will be split based on the size of input.

Dispatch requires a sort of $O(S\log S)$, a bin-count of $O(S)$, and an indexing operation of $O(SD)$. The overall complexity $O(SD + S\log S)$ is much smaller than the batch matrix multiplication of size $O(S^2EDC)$. The additional cost is modest and only an extra all-to-all collective whose message size is minimal. Our dynamic gating ensures token dropping will not happen, improving the model's robustness. 
Our dynamic gating also ensures that no empty placeholders will be transferred between devices, removing the waste in memory allocation of the reordered input and communication volume.

After all tokens are processed by their assigned experts, they are collected through another all-to-all communication collective, sent to their original device, and restored to their original order. 
This is typically implemented using batch matrix multiplication (BMM) but, as in the first stage, BMM can be replaced with an indexing operation that reduces complexity.


\subsection{Reduced Latency and Memory Usage}
\begin{figure*}[t]
\centering
\includegraphics[width=.85\textwidth]{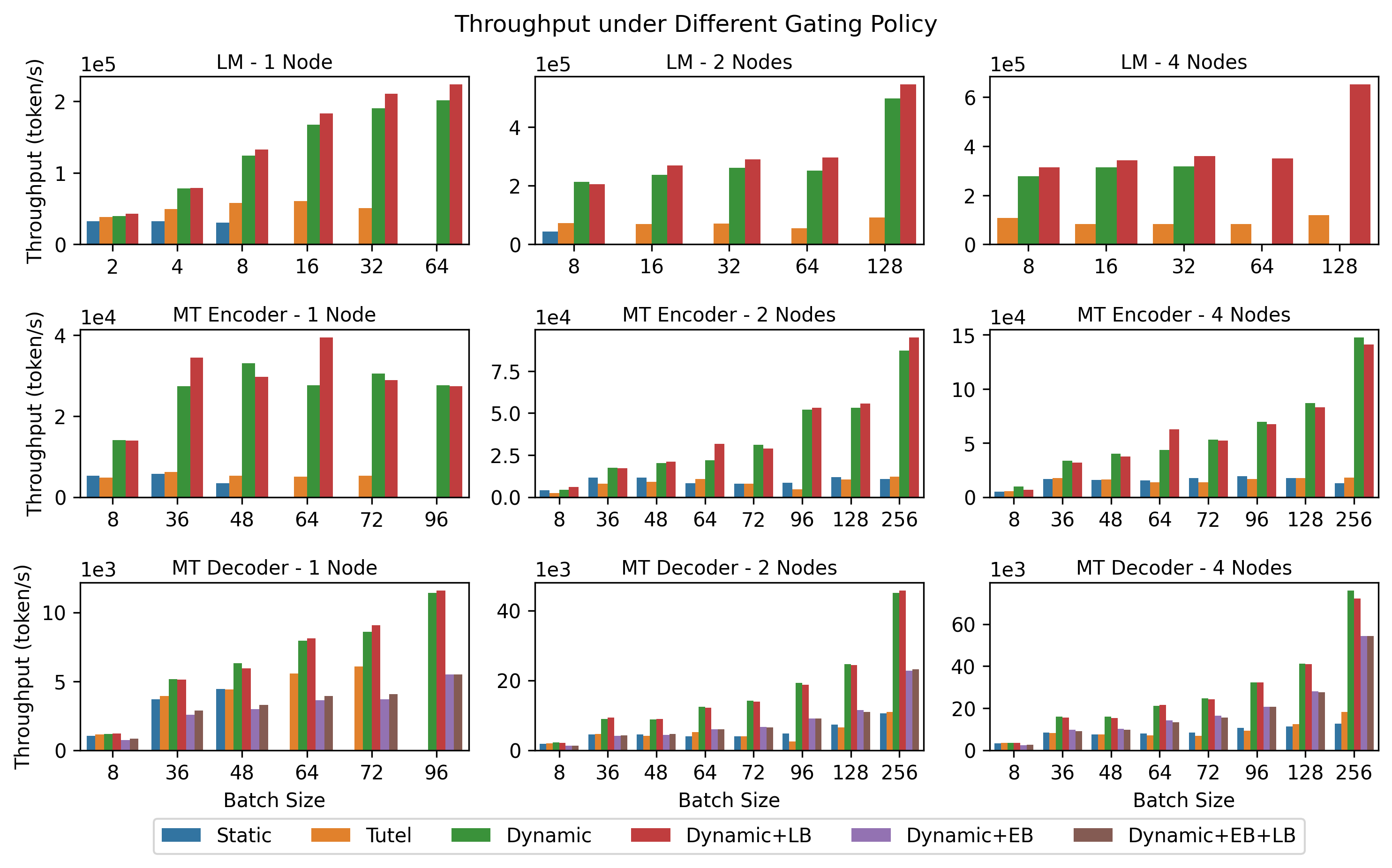}
\caption{{Throughput comparison of different gating policies in MoE models, including static gating (baseline), Tutel gating\cite{tutel}, our dynamic gating policy (Sec.~\ref{sec:dynamicgate}), dynamic gating with Load Balancing (LB, Sec.~\ref{sec:loadbal}), dynamic gating with Expert Buffering (EB, Sec.~\ref{sec:expcache}), and all optimizations combined. Missing bars represent infeasible cases under the corresponding policy and batch size. Eg. Tutel cannot support beyond batch size=32 for LM-1 Node.
Dynamic gating reduces memory usage and message sizes in communication, enables larger batch sizes and substantially faster processing times than static gating. Expert buffering trades latency for smaller memory usage while still achieving higher throughput on the MT Decoder. Load balancing further improves latency when combined with dynamic gating and expert buffering. Note that load balancing only makes sense in the context of dynamic gating where each expert gets different number of tokens. Load balancing particularly shines under multi-node setting or combined with expert buffering as it can improve cache miss rate (See Fig~\ref{fig:loadfix}).}}
\label{fig:dynamicspeedlm}
\end{figure*}

\begin{figure}[t]
\centering
\includegraphics[width=.45\textwidth]{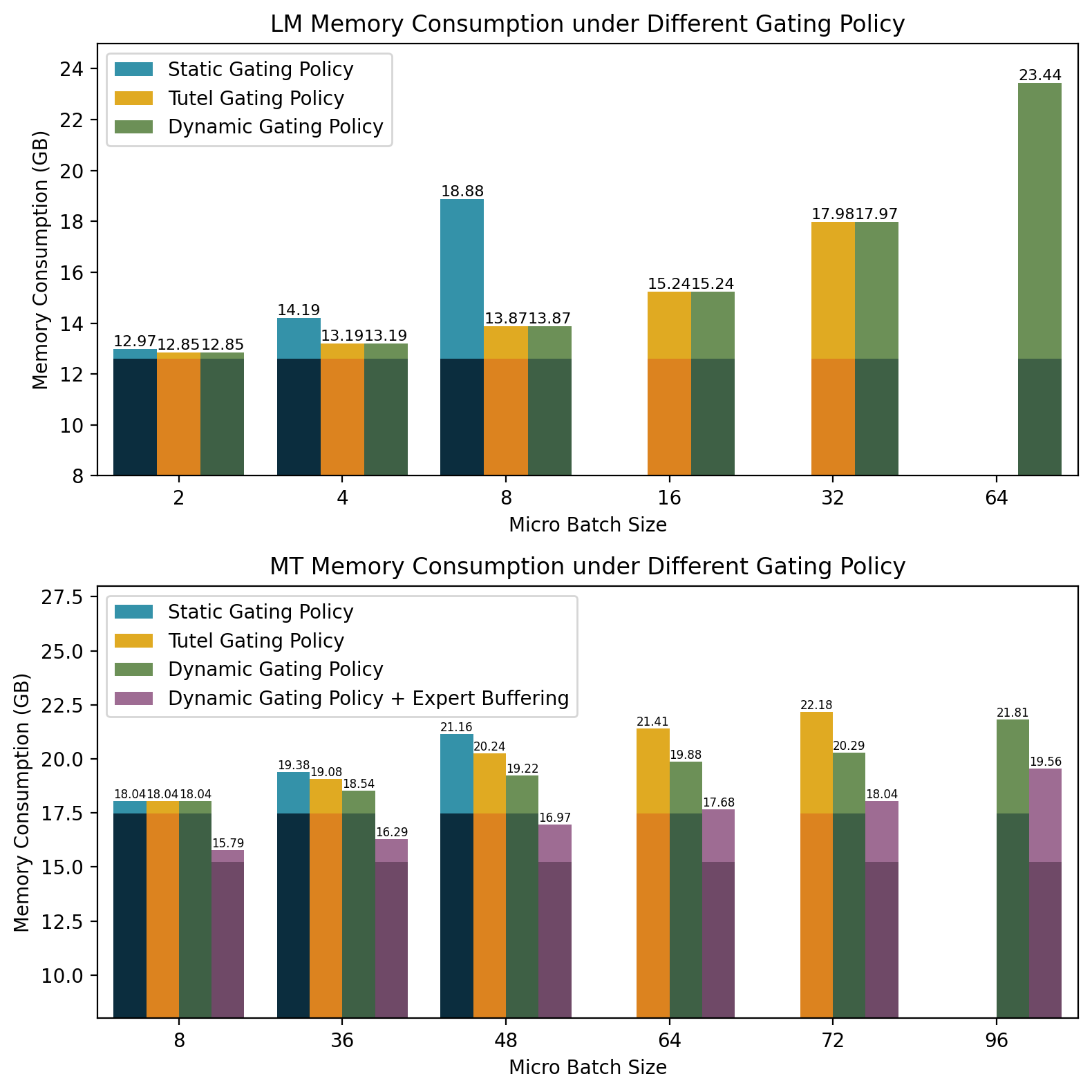}
\caption{{Comparison of memory consumption between MoE models under static and dynamic gating policy. Light shade represents dynamic memory allocation (activation memory). Dark shade represents static memory allocation (model parameters). Missing bars in each plot capture the infeasible cases under the corresponding policy.
Compared to Static and Tutel\cite{tutel}, Dynamic Gating reduces the memory usage, thus enabling larger batch sizes. Expert Buffering further reduces the memory consumption of model parameters.}}
\label{fig:dynamicmem}
\end{figure}

We study the impact of the dynamic gating policy on execution time and memory usage on the LM and MT MoE use cases. 
We keep the machine configuration same as before for a fair comparison.
As the dynamic gating policy introduces workload imbalance between different GPUs, for each case, we study the impact of different datasets and tasks on the performance of the model. The experiment is executed by forwarding ten independent epochs on each subset/subtask and recording the average throughput and memory consumption for each batch across the experiments.

Figure~\ref{fig:dynamicspeedlm} compares the impact of different gating policies on throughput. Results on Expert Buffering and Load Balancing will be explained in Section~\ref{sec:expcache} and~\ref{sec:loadbal}, respectively.
Our dynamic gating policy significantly increases throughput across all batch sizes and tasks, compared against the baseline and Tutel gating\cite{tutel}. By removing the large dispatch mask, dynamic gating also enables larger batchsizes under the same amount of resources. This improves throughput by up to $6.21\times$/$3.32\times$ when compared against static/Tutel gating under single-node LM, by up to $5.75\times$/$5.33\times$ under MT Encoder, and by up to $2.58\times$/$1.88\times$ under MT decoder.

Since dynamic gating removes the waste factor by reducing the volume of communication, it also reduces communication overheads and improves throughput. Therefore, the benefit of the dynamic gating policy widens when the model is deployed on multiple nodes, where communication overheads are more prominent. Dynamic gating improves throughput by up to 11.55$\times$, 10.98$\times$ , 5.71$\times$ in LM, MT Encoder, and MT decoder, respectively, when compared against static gating.



Figure~\ref{fig:dynamicmem} summarizes the effect of different gating policies on memory consumption, using single node cases as an example. Dynamic gating enables larger batch sizes, which runs even faster, when compared against the static gating policy with smaller batch sizes. Dynamic gating reduces the memory footprint by removing the dispatch mask, and also reduces wasted memory allocations for empty paddings and placeholders. As a result, the  memory allocated for the activation (bright colors in the figure) for LM with batch size of 8 falls from 6.29GB to 1.28 GB, which is a 79.6\% decrease. For MT with batchsize of 8, memory allocations fall from 1.89GB to 1.05 GB, which is a 44.2\% decrease.
Reducing memory consumption also allows a larger batch size to be allocated under the same machine configuration. The dynamic gating version permits a batch size of 64 for LM and 96 for MT, which is 8$\times$ and 2$\times$ larger than batch size permitted on the static counterparts.
\section{Expert Buffering}
\label{sec:expcache}
\begin{figure}[t]
\centering
\includegraphics[width=.45\textwidth]{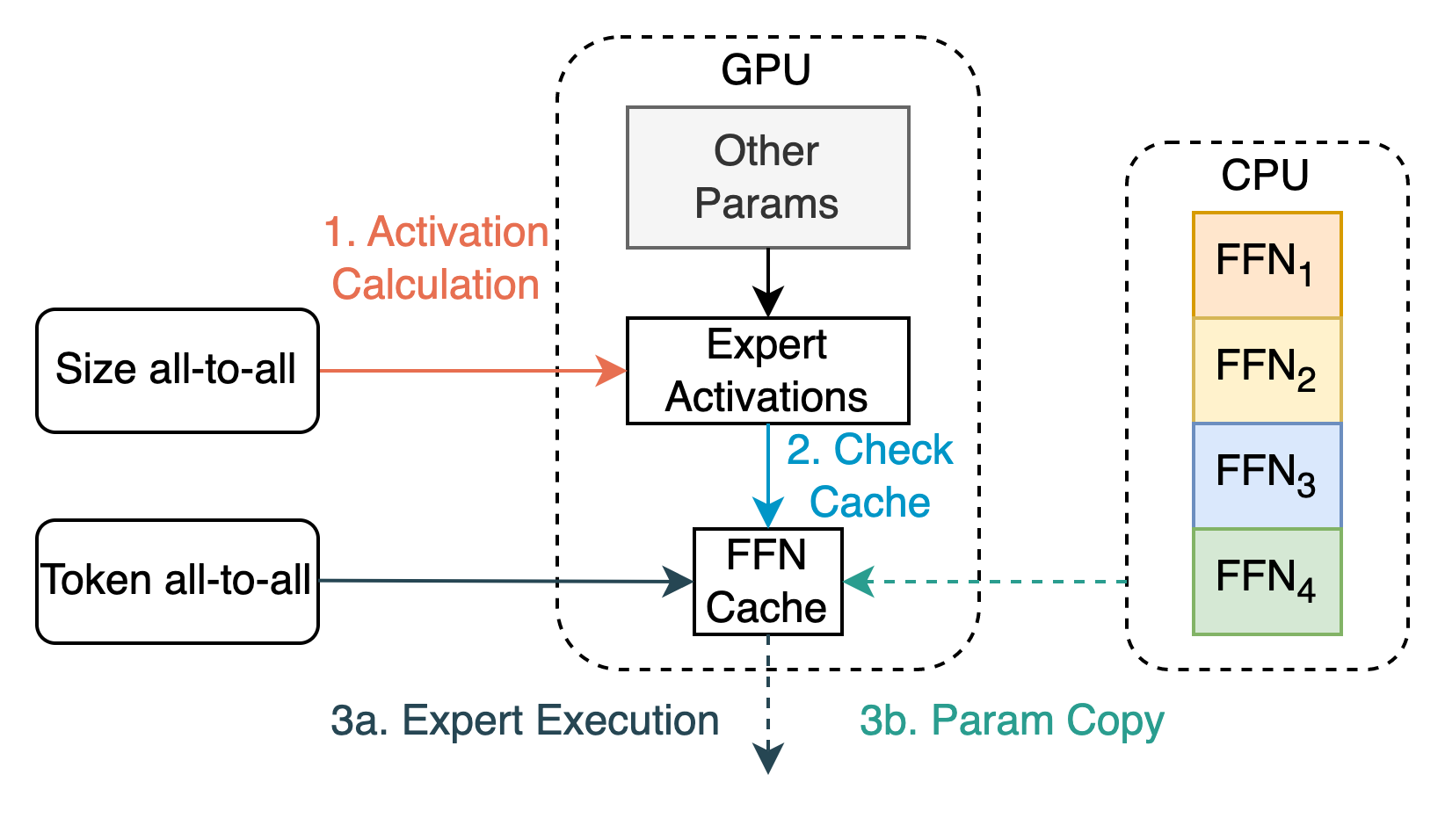}
\caption{Illustration of the Expert Buffering mechanism. We move the expert parameters to CPU memory to reduce burden on GPU memory. On GPU memory, we allocate space only for a few expert entries to buffer active or hot experts. (1) During inference, the all-to-all size message sent in stage 1 as shown in Figure~\ref{fig:dynamic}(b) signals which experts located in the current device are active. (2) Then the expert cache will check whether the active experts currently reside in the buffer. (3a) If found (cache hit), parameters in the expert buffer will be used to process the tokens. (3b) If not found (cache miss), then the expert parameters will be requested from the CPU memory. The number of cache entries on GPU memory is a tunable parameter to adjust for desirable GPU memory usage and latency (See Section~\ref{sec:expcache}).}
\label{fig:cache}
\end{figure}
Although dynamic gating reduces waste in computation and dynamic memory allocation associated with the gating function, static memory usage associated with the large number of MoE parameters still puts a huge burden on GPU memory at deployment.
The high sparsity in expert activation pattern prompts us to investigate whether there is a way to reduce the memory usage by pruning out the idle experts.

\subsection{Sparse Expert Activation}

Our investigation in the expert activation pattern shows that although, in every batch, there exists some experts that are inactive, all experts have been activated a few times across time and batches. Pruning out experts that are not frequently active can potentially hurt model accuracy. However, we can offload the less frequently accessed experts to CPU memory and use the GPU memory for hot and active experts. 


\begin{figure}[tbp]
\centering
\includegraphics[width=.40\textwidth]{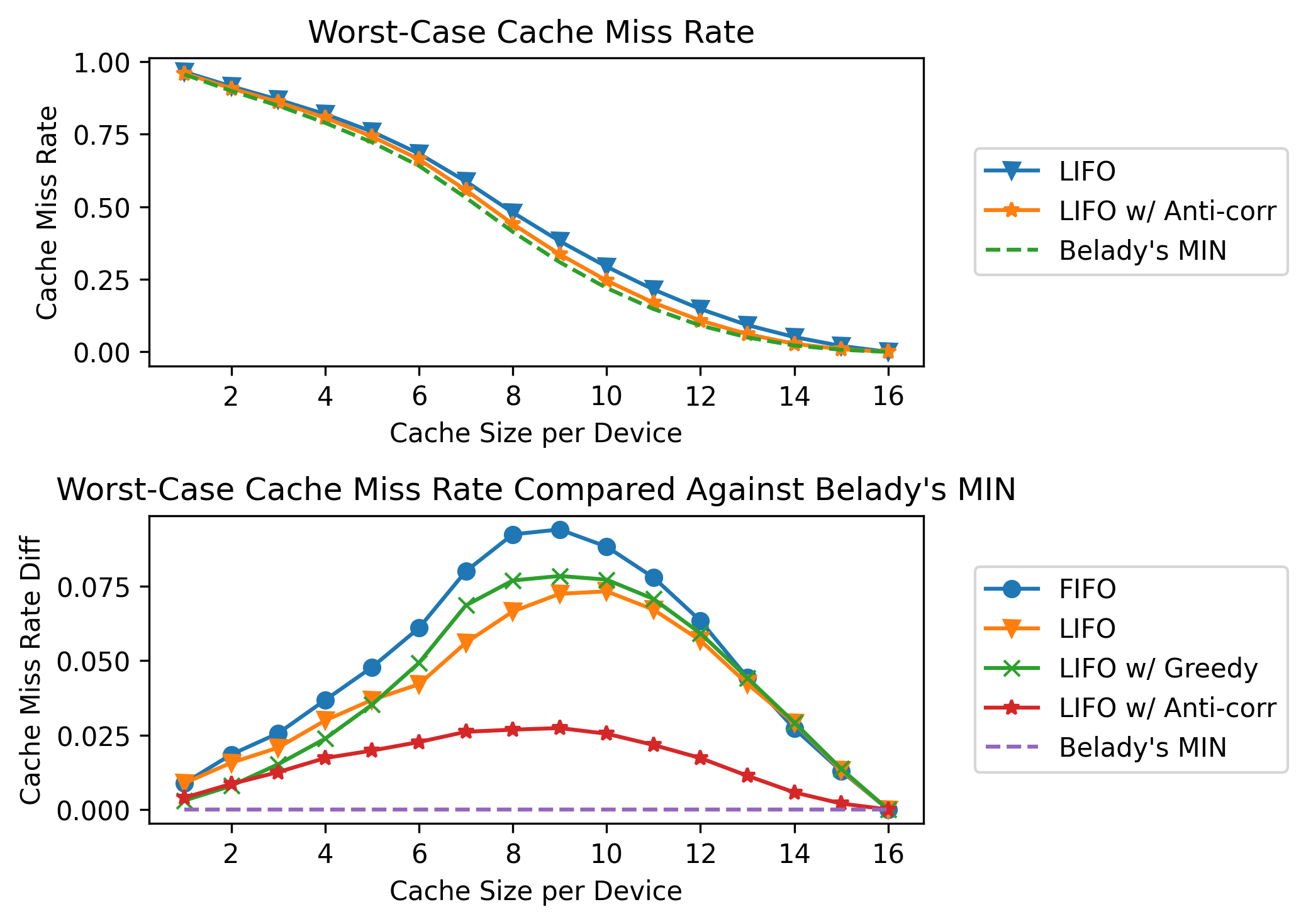}
\caption{Worst-Case Cache Miss Rate  obtained from traces of expert activations from MT decoders. 
We tune the cache size per GPU from 1 to 16, and examine the impact of cache size on the worst case Cache Miss Rate on different layers of MT decoders over different tasks. \textbf{(a)} caching performance with/without any reassignment. {\textbf{(b)} caching performance against the theoretical optimal Belady's MIN. The miss rate is further reduced by load balancing, and is very close to Belady's MIN (See Sec.~\ref{sec:loadbal}).}}
\label{fig:cache_miss}
\end{figure}

We propose the expert buffering mechanism to exploit expert sparsity and reduce static memory allocation. Figure~\ref{fig:cache} illustrates the mechanism, which reduces static memory consumption by offloading expert parameters to CPU memory. 
Since CPU is much slower than GPU for matrix multiplication, we only use CPU memory to hold the experts but do not offload the computation. We use GPU memory to cache active experts and perform computation. 

\subsection{Cache Management} 

During inference, under dynamic gating, once the gating decision is made by the gating function, each GPU receives the number of tokens assigned to its experts. If an expert receives a positive token count, it is considered active for the current batch. The process then checks if the active expert is already cached in GPU memory. If not, then the process will launch a Memcopy to transfer the required expert parameters into the cache. Copying expert parameters from CPU memory to GPU DRAM will be launched in parallel with all-to-all communication, to allow for overlap of data transfers and latency hiding. 

In cases where the cache is already full but more experts are needed, eviction will be triggered to make space for the new experts.
The eviction policy is designed as follows. First, we will first evict experts that are not active in this batch since they are also less likely to be used in the future due to temporal locality. Next, we will evict expert parameters under a Last In, First Out (LIFO) policy. 

The reason for adopting a LIFO policy is rooted in the implementation of recent MoE Transformers. If multiple experts are allocated to a single GPU, MoE Transformer will execute the experts serially in the increasing order of their ids.
Consider a small example of $E=4$ experts and cache size of $2$ experts, and assume expert (1, 2, 3) are needed. After stage 1, expert 1 and 2 will be pushed into the cache, and we need to evict one of them to load expert 3. By evicting expert 2 instead of 1, we ensure the expert with the shortest reuse distance is kept in the cache.



\subsection{Cache Miss Rates} 

To estimate the technical feasibility of Expert Buffering, we calculate cache miss rate for machine translation use case on layers that exhibit enormous expert sparsity.
We note that the cache is deployed per device, and each device caches experts that have been assigned to it. As a result, we may vary cache size from 1 to 16 experts, generating a saving of 0-32.2\% on total static memory allocation.
We calculate the global cache miss rate under each circumstance.
The {worst-case cache miss rates} are shown in Figure~\ref{fig:cache_miss}. We notice the Cache Miss Rate starts to decrease faster when the cache size is larger than 5 per GPU, which is a cache size of {40} in total. This result is consistent with our previous observation that there will be more than 90 experts being empty inside the decoder.
We further compare our caching policy with Belady's MIN, the theoretical optimal policy that requires information from the future. Fig.~\ref{fig:cache_miss}(b) shows that the LIFO policy is better than FIFO, and with the load balancing introduced in Sec.~\ref{sec:loadbal}, our policy can obtain a cache miss rate very close to Belady's MIN.

To evaluate the impact of the proposed mechanism, we perform experiments on Expert Buffering on the MT Decoder. 
The cache size is selected to be around 80 experts in total, which fits 10 experts per GPU under single node case. This is the point where the cache miss rate starts showing saturation behavior in Figure~\ref{fig:cache_miss}(a).

{Figure~\ref{fig:dynamicspeedlm} and Figure~\ref{fig:dynamicmem}} show the impact of expert buffering on throughput and the static memory allocations for MT. Expert buffering has successfully reduced the static memory consumption by 2.25GB. This memory reduction is particularly useful for users with limited number of GPUs. 
Although the throughput becomes smaller compared to memory-intensive dynamic gating, the throughput obtained by expert caching is comparable to baselines under single-node, and 2.21$\times$/4.30$\times$ under 2/4 nodes. 



\begin{figure}[tbp]
\centering
\includegraphics[width=.40\textwidth]{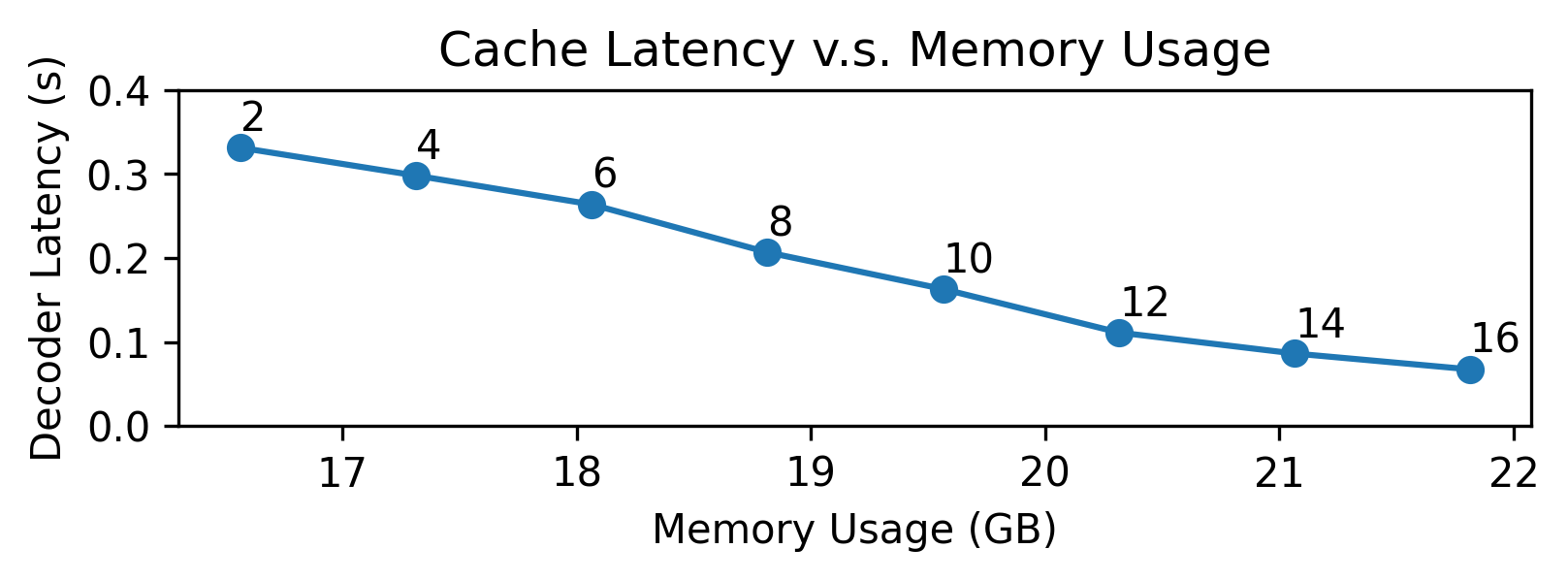}
\caption{Tradeoff between memory and latency under different cache configuration on MT Decoder. 
Corresponding cache size per GPU is marked on the plot.}
\label{fig:cache_memtime}
\end{figure}

Furthermore, we study the latency-memory tradeoff incurred by the expert caching mechanism, and estimate the pareto frontier. Figure~\ref{fig:cache_memtime} shows the latency and memory consumption under a series of cache configurations. We vary the cache per GPU and measure the decoder latency and peak memory consumption. The result shows that the pareto frontier is similar to the outliers on the cache miss rate plot in Figure~\ref{fig:cache_miss}(a). 

Our findings also indicate that the primary contributor to increased latency is the constrained CPU-GPU bandwidth, which we observed to saturate at 12GB/s during our experiments. 
The findings indicate that layers with a high cache miss rate may impede performance. Adopting  technologies that enhance CPU-GPU bandwidth, such as the NVIDIA Grace Hopper superchip and PCIe v5 can mitigate the latency issues in situations where memory is constrained.

We note that no prior work exploits the unique characteristics of MoE Transformers to optimize memory usage. As a caching strategy that is specifically tailored for MoE models, expert buffering is orthogonal to prior memory saving mechanisms such as offloading~\cite{ren2021zero,shen2022se} and can be seamlessly integrated for greater memory savings.
\section{Load Balancing}
\label{sec:loadbal}

As we saw in Section~\ref{sec:expass}, token assignments to experts are highly imbalanced, hence the load assigned to each device is also highly imbalanced. Those devices hosting hot experts can become bottlenecks and become more vulnerable to out of memory error. Moreover, devices hosting cold experts may sit idle while waiting for devices that are hosting hot experts to finish their load. As a result, load balancing is critical for having a robust and stable model. 

We propose a simple load balancing scheme during the model deployment. We optimize the allocation of experts by leveraging historical load data. Specifically, we encode historical expert allocation into a matrix, and balance the load on each device accordingly.
We combine higher-loaded experts with lower-loaded experts, so that the load can be distributed to different devices. 
We denote the expert placement with $P_{mn}$, where m is the expert id ($m=1\ldots E$), n is the device id ($n=1\ldots D$) and $P_{mn} = 1$ indicates that the $m$-th expert is allocated on the $n$-th device. We also denote the expert activation with $A_{mb}$, where m is the expert id and $b$ is the batch id ($b=1\ldots B$), and $A_{mb}$ represents the fraction of tokens assigned to expert $e$ at batch id $b$.
The problem can be thus formalized as follows:
$$ \min \max_{m, b} |\sum_{n} P_{mn} A_{mb} - \frac{1}{D}| \text{ subject to } \sum_m P_{mn} = \frac{E}{D} \ \forall n$$
This problem can be reduced to the multi-way number partitioning problem~\cite{graham1969bounds}, which is NP-hard. To balance the memory usage and simplify the communication process, each GPU should be assigned the same number of experts.

\subsection{Greedy Balancing for Independent Activation}

We utilize a greedy algorithm to generate approximations to the optimal assignment. We sort the experts by their average work load in historical data $\tilde{A}_{m}$, and assign the experts to GPUs on a descending order. At each step, an expert is assigned to the GPU with the smallest load, calculated by $\sum_{m}P_{mn}\tilde{A}_{m}$. 
Once a GPU reaches the designated capacity, it will be removed from the list of candidates.
\begin{figure}[t]
\centering
\includegraphics[width=.48\textwidth]{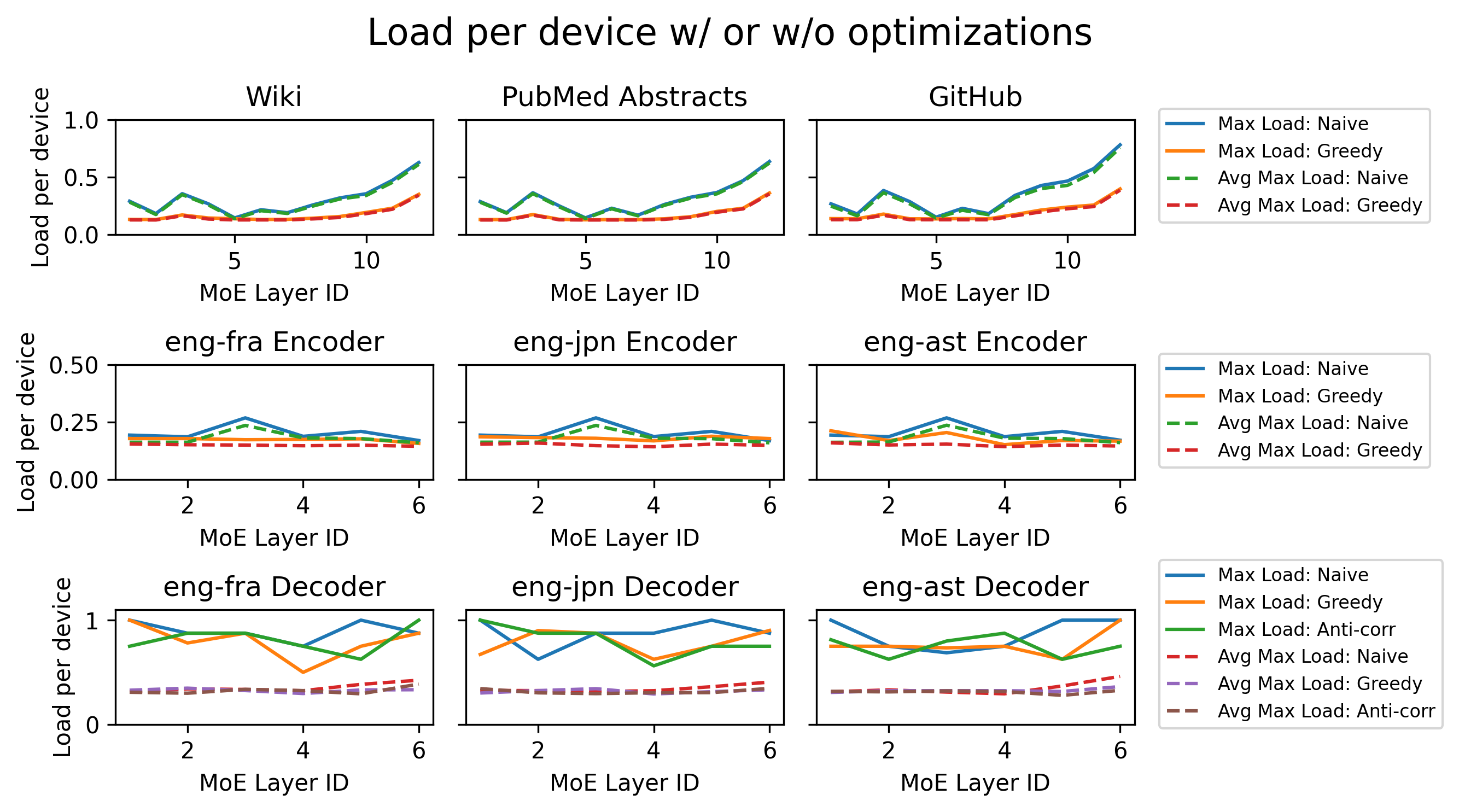}
\caption{The effect of our proposed load balancing mechanism: Based on historical activation data, our algorithm is able to significantly reduce the load imbalance problem on LM tasks and improve the robustness by limiting the maximum workload on a single device.}
\label{fig:loadfix}
\vspace{-0.1in}
\end{figure}



Figure~\ref{fig:loadfix} summarizes the balance of the load under the original order and the new order. The balance of load is estimated using existing activation data introduced in Section~\ref{sec:dynamic}. 
To perform the experiment, we separate the data into two halves. We use the first half of the activation data to generate a device assignment for each expert, then estimate the work load under generated assignment using the second half of the activation data. Results are normalized by the total batch size, which means the numbers represent the share of the total number of tokens each device will handle in a certain batch. 

We record the Max Load, which is the maximum share of the load that has ever appeared in all batches on the test set, and the Avg Max Load, which is the maximum share of the load averaged over all batches. The Max Load estimates the worst case scenario that relates to out-of-memory error, and the Avg Max Load estimates the average case where imbalance on work load can lead to bottlenecks and harm inference speed.

Results show that Greedy is able to balance the load for LM use case by improving both metrics (Max Load and Avg Max load per device) significantly, reducing from more than 0.6 to less than 0.4. Similarly for Machine Translation use case, Greedy can balance the expert load assignment for the MT encoder. 
Figure~\ref{fig:dynamicspeedlm} shows the benefit of the Greedy Rebalancing on the LM and MT Encoder: Rebalancing increases the throughput by a maximum of 10.1\% and 19.5\% compared against pure dynamic gating. Furthermore, rebalancing allows the LM to achieve a batchsize of 64 and 128 when it is deployed on four nodes, making the model more robust.

\subsection{{Anti-correlation Balancing for Correlated Activation}}

Greedy is less effective for the Machine Translation Decoder. We found that expert activation level becomes a less effective indicator in this case due to correlation between experts. To handle this problem, we propose Anti-correlation Balancing, which takes correlation into consideration. Denoting the {Pearson} correlation between the current expert $a$ to expert $b$ in the {historical data} as $S_{ab}$, the current work load can be modified from $\sum_{m}P_{mn}\tilde{A}_{m}$ to $\sum_{m}P_{mn}(\tilde{A}_{m} + 0.5 * S_{am})$. This algorithm successfully reduces the Avg Max Load and the Max Load on most cases. We notice that a more balanced work load also has a positive impact on the cache miss rate. As shown in Figure~\ref{fig:cache_miss} and~\ref{fig:dynamicspeedlm}, the worst-case cache miss rate decreases for all cache sizes over MT Decoders, which leads to a maximum increase of 1.9\% on their throughputs. 
\section{Related Work}
While the MoE Transformer substantially reduces the training cost and FLOPS for large models, the outrageous size of MoE Transformers and the complex expert parallelism \cite{lepikhin2020gshard} poses obstacles for its deployment, including the high GPU memory requirement and the excessive communication overhead of expert assignment. Various approaches have been invented to relieve these obstacles. Switch Transformer \cite{fedus2021switch} and ELSLM \cite{artetxe2021efficient} use knowledge distillation to distill a large MoE Transformer into a dense model. While distillation reduces the number of parameters, only a small portion (about 30\%) of the accuracy gain can be retained. {The MoS strategy proposed in DeepSpeed-MoE \cite{rajbhandari2022deepspeed} distills the knowledge to a smaller MoE Transformer with less layers and shared experts.} SE-MoE \cite{shen2022se} uses pruning to reduce the number of experts in the model. 
WideNet \cite{xue2021go} and MPoE \cite{gao2022parameter} reduce the number of parameters by enforcing parameter sharing. Beyond reducing the parameters, other methods directly reduce computation and communication. The BASE Layer and Switch Transformer also reduce the number of experts each token is assigned to reduce the communication volume and computation. V-MoE \cite{riquelme2021scaling} further reduces the number by dropping out a large portion of tokens. Hash Layer \cite{roller2021hash} replaces the gating layer with a precomputed hash function, which reduces the computation cost, but doesn’t alleviate the communication overhead. 
As the MoE Transformer is a type of Transformer, techniques and optimized architectures that enhance Transformer inference speed may apply. Relevant examples include Reformer\cite{kitaev2020reformer}, Longformer\cite{beltagy2020longformer}, and Terraformer\cite{jaszczur2021sparse}. However, there is scant discussion of their application to MoE Transformers.


Offloading and swapping strategies such as \cite{huang2020swapadvisor} swaps unused tensors form the GPU memory to the main memory to reduce the resource requirement. However, existing strategies can only be applied on dense models. Applying these strategies efficiently on conditional neural networks such as MoE is non-trivial, since the data flow graph cannot be constructed in advance due to the conditional computation. FastMoE \cite{he2021fastmoe, he2022fastermoe} designed customized communication primitives and gating kernels for token assignment to reduce the communication overhead, but it has not been tested on outrageously large neural networks. 
Tutel\cite{tutel} and DeepSpeed-MoE \cite{rajbhandari2022deepspeed} improve MoE model performance on datacenter-scale systems by combining system and architecture methods with tailored kernels for both Transformer and MoE layers, and specialized communication primitives. The approach combines expert parallelism, model parallelism, and tensor parallelism to significantly boost throughput and reduce latency. However, DeepSpeed-MoE is not designed to conserve GPU resources and therefore may be impractical for many academic users. SE-MoE\cite{shen2022se} utilizes Ring Memory offloading to reduce GPU usage, achieving better throughput than DeepSpeed-MoE in low-resource scenarios. However, this approach does not leverage expert parallelism from MoE Transformers.

\section{Conclusion}
\label{sec:conclusion}
While at training time, mixtures of expert (MoE) models show superior performance to their flop-equivalent dense counterpart models, they are notoriously large, need a large number of GPUs to deploy and hard to democratize. Researchers outside large industry labs do not have access to hundreds or thousands of GPUs to afford exploring such large models. Moreover, they are much slower than their dense counterparts at inference.  To overcome these challenges, we propose three optimization techniques (Dynamic Gating, Expert Buffering, and Expert Load Balancing) to improve memory and latency profile of such models for deployment.




\end{document}